\pdfoutput=1

\documentclass[twocolumn,amsmath,amssymb,superscriptaddress,prb]{revtex4}
\usepackage{graphicx}
\usepackage{bm}
\usepackage{subfigure}
\usepackage{multirow}
\usepackage{color}

\begin{document}


\title{Proposal for a Graphene Plasmonic THz Emitter}

\author{Don C. Schmadel}
    \affiliation{Department of Physics, University of Maryland at College park, College Park, Maryland, 20742, USA}
    \affiliation{Center for Nanophysics and Advanced Materials, University of Maryland at College park, College Park, Maryland, 20742, USA}

\author{Gregory S. Jenkins}
    \homepage{http://www.irhall.umd.edu}
    \email{GregJenkins@MyFastMail.com}
    \affiliation{Department of Physics, University of Maryland at College park, College Park, Maryland, 20742, USA}
    \affiliation{Center for Nanophysics and Advanced Materials, University of Maryland at College park, College Park, Maryland, 20742, USA}

\author{H. Dennis Drew}
    \affiliation{Department of Physics, University of Maryland at College park, College Park, Maryland, 20742, USA}
    \affiliation{Center for Nanophysics and Advanced Materials, University of Maryland at College park, College Park, Maryland, 20742, USA}
\date{\today}

\begin{abstract}
We propose a terahertz radiation source  based on the excitation of plasma resonances in graphene structures by means of mixing two NIR laser signals with a THz difference frequency.  The process is the photo-thermo-electric effect which has recently been demonstrated to be operative at THz frequencies in graphene.  An antenna couples the THz radiation out of the sub-wavelength graphene element and into the far field.  The emission is monochromatic with a bandwidth determined by that of the NIR laser sources.  The output power of the device as a function of the emitter frequency is estimated at $10$'s of $\mu W$'s.
\end{abstract}

\maketitle

\section*{Introduction}
Terahertz (THz) radiation has important uses from security to medicine,\cite{1-cai2013} however the THz spectral range is notoriously underdeveloped because of the lack of room temperature sources and detectors.  As a result little technology exists for this otherwise attractive ultra high frequency band. Recent work suggests that graphene, a single atom thick carbon honeycomb lattice, may be the key to a robust THz technology.  Graphene has a gapless Dirac electronic spectrum and a strong specific coupling to light.  The gapless character implies optical coupling at all frequencies.  Moreover, doped graphene exhibits plasmonic charge density waves that allow resonant excitation in micron sized samples with dipole radiation.\cite{ju_2011} Recently, room temperature THz detectors have been demonstrated that operate on a photo-thermo-electric principle.\cite{1-cai2013}  THz absorption in a graphene element raises the temperature of the graphene carriers, which then diffuse to the contacts made of dissimilar metals and produces a photo voltage proportional to the Seebeck coefficient of the graphene.

In this paper we propose a room temperature tunable THz emitter based on the photo-thermo-electric effect in graphene. There are several sources for THz radiation currently available, but they all have significant limitations or disadvantages. For example both photo-mixing and direct mixing in GaAs offer only $\sim 1 \,\mathrm{mW}$ up to $1 \,\mathrm{THz}$.\cite{3-brown1995,blaney1974} The same is true for backward wave oscilators.\cite{korolev2001}
The quantum cascade laser \cite{faist1994} offers significant power and tunability but only at high THz frequencies and at a very high cost. We are proposing a new kind of photo-mixing in single layer graphene where near-infrared (NIR) laser radiation generates photo-thermal voltages that excite plasmons in a graphene element that then radiate their energy to the far field by means of an antenna.  Some simple implementations of this approach are described below followed by a brief discussion of various physical processes that affect the operation.

\section*{Photo-thermal-electric plasmonic emitters}
The carriers in doped graphene form a two-dimensional (2D) electron gas, whose dynamics at femtosecond time scales are essentially decoupled from the supporting lattice. Absorption of a photon in the NIR or visible range will excite an electron/hole pair producing carriers far from the Fermi level.  Subsequent electron/electron scattering, occurring on a time scale of tens of femtoseconds,\cite{2-hwang2007} quickly distribute the absorbed energy of the electron with other carriers resulting in a local quasi-thermal distribution. Diffusive transport of these local hot carriers will occur up to frequencies corresponding to these time scales, in particular at THz frequencies.  A photon source for this type of excitation could be a spatially and/or temporally patterned NIR laser beam.  When directed onto a doped graphene sheet it would create a spatial thermal pattern of localized hot carriers within the graphene electron gas. The Seebeck effect will produce an electric field corresponding to the gradient of the thermal pattern. Temporally varying the NIR pattern at THz frequencies will result in THz currents. Such temporal variation can be attained by interfering two separate NIR beams whose difference frequency is in the THz spectral region. Generation of plasmons in the graphene sheet will occur if the spatial and temporal features of the incident radiation intensity are related to the plasmon dispersion:
\begin{equation*}\label{PlasmaFreq}
\omega_p^2=\frac{e^2 E_F}{\epsilon \hbar^2 \lambda}
\end{equation*}
where $\lambda$ is the plasmon wavelength, $\omega_p = 2 \pi f$ is radial frequency of the plasmon which corresponds to the difference frequency of the incident NIR light, and $\epsilon$ is the dielectric constant of the environment of  the graphene sheet.

\begin{figure}
\includegraphics[scale=.5]{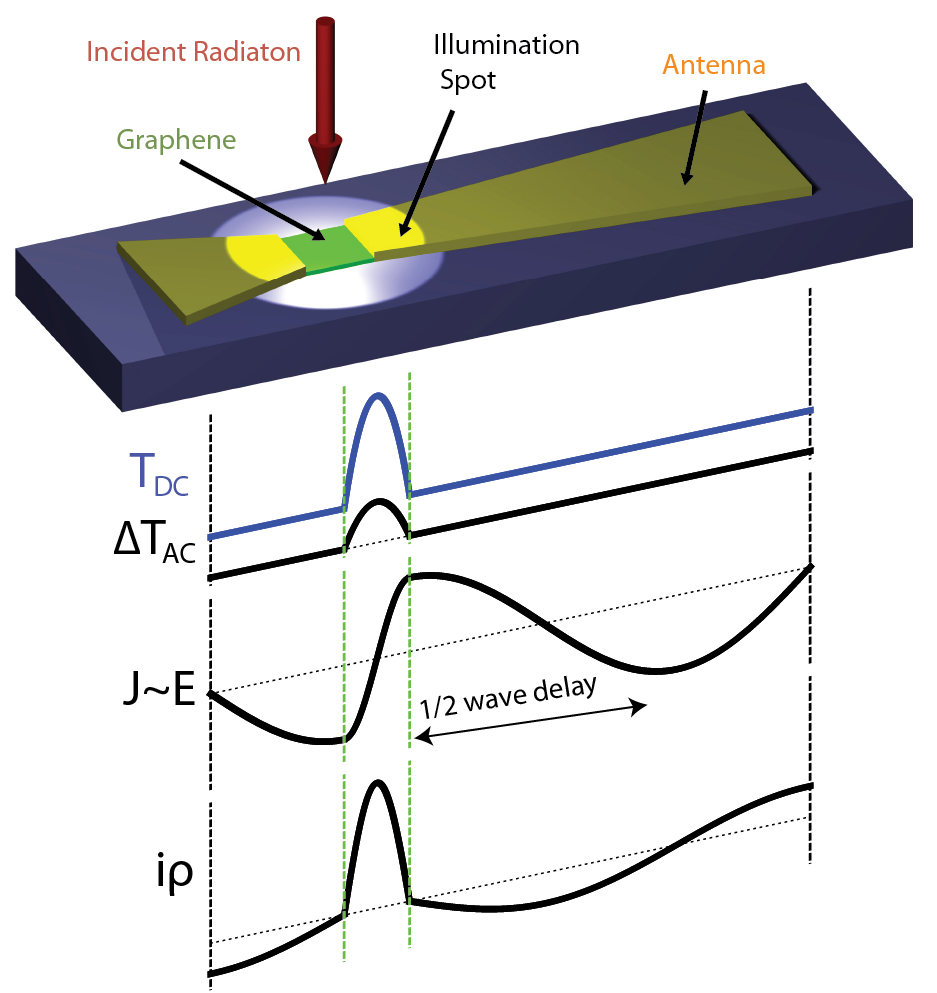}
\caption{\label{fig1}(color online) A Graphene strip of length L couples to an asymmetric antenna.  Uniform illumination by two interfering collinear laser beams with a THz difference frequency excites a standing plasma resonance in the graphene through the photo-thermo-electric effect.  The sub-wavelength energy in the graphene is coupled into the far field  using an antenna.  The graphs schematically illustrate the resulting temperature profile, and the currents, fields, and charge densities in the graphene and the antenna. The monopolar plasma excitation for this case requires an asymmetric antenna for efficient coupling.}
\end{figure}

Figure \ref{fig1} illustrates a particular configuration consisting of a graphene strip whose length (the distance between the antenna contacts) is chosen to have a plasma resonance corresponding to the desired THz frequency.  The graphene element is coupled to an antenna for far field emission.  Two incident interfering collinear NIR laser beams of different optical frequencies directed as a spot onto the graphene will heat the carriers in the graphene strip and, through the thermo-electric effect, excite a plasma resonance whose frequency corresponds to their difference frequency.  The low wave impedance of the graphene relative to the metal of the antenna results in antinodal current boundary conditions for the graphene plasmons.

In the structure illustrated in Fig. \ref{fig1}, uniform illumination by the NIR laser beams produces a symmetric temperature distribution which is clamped at $T=T_0$ by the metal contacts.  The plasma excitation excited by this temperature distribution will have an even charge distribution but antisymmetric currents and electric fields. Therefore, the resonance represents a monopole oscillation.  The standing plasma oscillation may be considered as two running plasmons traveling in opposite directions each of which will excite currents in the antenna. However, because these currents are of opposite phase, the antenna must be asymmetric in order to radiate efficiently, as will be discussed later. As described above the input laser beam delivers energy to  the electron system that is converted to heat by electron/electron scattering. We may represent the result as a simple heat input flux with spatial and temporal variation, $I(x,t)=I(1+e^{-i \omega t})$, where $\omega=\omega_1 - \omega_2$ is the THz difference frequency.

The photo-thermo-electric response to this heat input is diminished by the cooling of the electrons through heat transport to the lattice.  However, diffusion cooling dominates provided that the sample length is less than the diffusion length $\xi = \sqrt{k_e/G}$ where $G$ is the thermal conductance to the lattice and $k_e = \Lambda \sigma T$ is the electron thermal conductance with $\Lambda = \frac{\pi^2}{3}(\frac{k_B}{e})^2 $ is the Lorenz number and $\sigma$ is the electrical conductivity. The thermal conductance due to acoustic phonon emission has been calculated and leads to diffusion lengths of several microns for typical doped graphene samples.\cite{6-viljas2010} However, recent attention to this issue has focused on the enhancement of the lattice conduction from the increased phase space for phonon emission due to disorder.\cite{7-song2012} This leads to electron cooling to the lattice of the form $H_l=-A(T^3 - T_0^3)$.  Measurements support this effect\cite{8-graham2013} and give a cooling rate of $A/\alpha \approx 5.5 \times 10^8 \, \mathrm{K}^{-1} \mathrm{s}^{-1}$ for a sample with $\mu \approx 8000 \, \mathrm{cm}^2 \mathrm{V}^{-1} \mathrm{s}^{-1}$ and $n\approx 10^{12} \, \mathrm{cm}^{-2}$, where $\alpha$ is the specific heat coefficient of the electron gas.  This implies a diffusion length $\xi\approx 1 \, \mu\mathrm{m}$ at room temperature. The emitters that we are considering are optimized for lengths less than $1 \, \mu\mathrm{m}$.  Therefore $H_l$ will be neglected in our analysis.

The electrical transport in response to the absorbed incident modulated NIR intensity, I(x,t),  is therefore governed  by the general transport equations and we can write:	
\begin{align}\label{TransportEq}
\begin{split}
  \nabla J &= \sigma \nabla E + e T \Lambda \sigma'(-\nabla^2 T) = \dot{\rho} \\
  \\
 \nabla J_u &= e T^2 \Lambda \sigma' \nabla E + e T \Lambda \sigma(-\nabla^2 T)\\
  &= I(x,t) + \frac{1}{2} \textrm{Re}(E \cdot J^\ast)-C_e \frac{dT}{dt}
\end{split}
\end{align}
where, $J$ is the electrical current density, $J_u$ the heat current density, $\rho$ the charge density, $C_e=\frac{\pi^2}{3}D(E_F)k_B^2T=\alpha T$ is the electron specific heat with $D(E_F)$ as the density of states at the Fermi energy $E_F$, $\sigma'= \frac{\partial \sigma}{\partial E} \big|_{E_F}$ and $\sigma$ is the electrical conductivity.  The $\frac{1}{2}\mathrm{Re}(E \cdot J^\ast)$ term represents the power transferred to the electron system by the decay of the plasma resonance. For the geometry of Figure \ref{fig1} with uniform intensity $I=I_0$, the spatial dependence of $\Delta T = T - T_0$, $\nabla J$, and $\nabla E$ are all even symmetric. The time average incident laser power produces a static temperature rise: $T_{DC}^2(x) - T_0^2=\frac{I_0}{\Lambda \sigma(0)} x (L-x)$.

The general solution to the transport equations given in Eq. \ref{TransportEq} can be found by expanding $ T - T_0$, $\nabla J$, and $\nabla E$ in a Fourier series.  We take the origin taken on one of the contacts so that $\Delta T = \sum_{n} A_n \sin{\frac{2 n \pi}{L}x}+B_n \cos{\frac{2 n \pi}{L}x}$. We will focus on the even $n=0$ mode which has the largest effect. The in-plane electric field and currents are related through the wave impedance of the plasmons:\cite{ 4-rana2008} $E=\frac{-i k}{2  \bar{\epsilon} \omega} J = -i k Z W J$, where $Z=(2 \bar{\epsilon} \omega W)^{-1}$ is the wave impedance, $W$ is the width of the graphene strip and $\bar{\epsilon}$ is the effective dielectric constant of the environment of the graphene.

For simplicity we consider these equations in the linearized form for $\Delta T = T(x) - T_0$ which is sufficient to illustrate the principle and allow estimations of the output power of the device.  We ignore the $E\cdot J^\ast$ term since it will be small compared with the total power, even at optimum operating conditions, due to the Carnot efficiency.  The solution for the current amplitude for the $n=0$ Fourier term is then
\begin{equation}\label{eqCurrentAmplitude1}
J_0=\frac{e T_0 \Lambda \sigma' k^2 I_0}{(i \omega C - \Lambda T_o \sigma k^2)(k+i \sigma k^2 Z)+ i e^2 T_0^2 \Lambda^2 \sigma'^2 k^4 Z}
\end{equation}
where $k=\pi/L$ and $I_0= 4 I /\pi$. The carrier scattering in graphene for the relevant conditions for this device is dominated by elastic screened-Coulomb scattering so that small angle scattering is not strong.  Therefore the relaxation time approximation is appropriate for both electrical and thermal transport coefficients and so the Drude model, $\sigma = n e^2 / m (\gamma - i \omega) = E_F e^2 / \pi \hbar^2 (\gamma - i \omega)$, may be used, where $\gamma \equiv 1/ \tau$ is the carrier scattering rate.

\begin{figure}
\includegraphics[scale=.5]{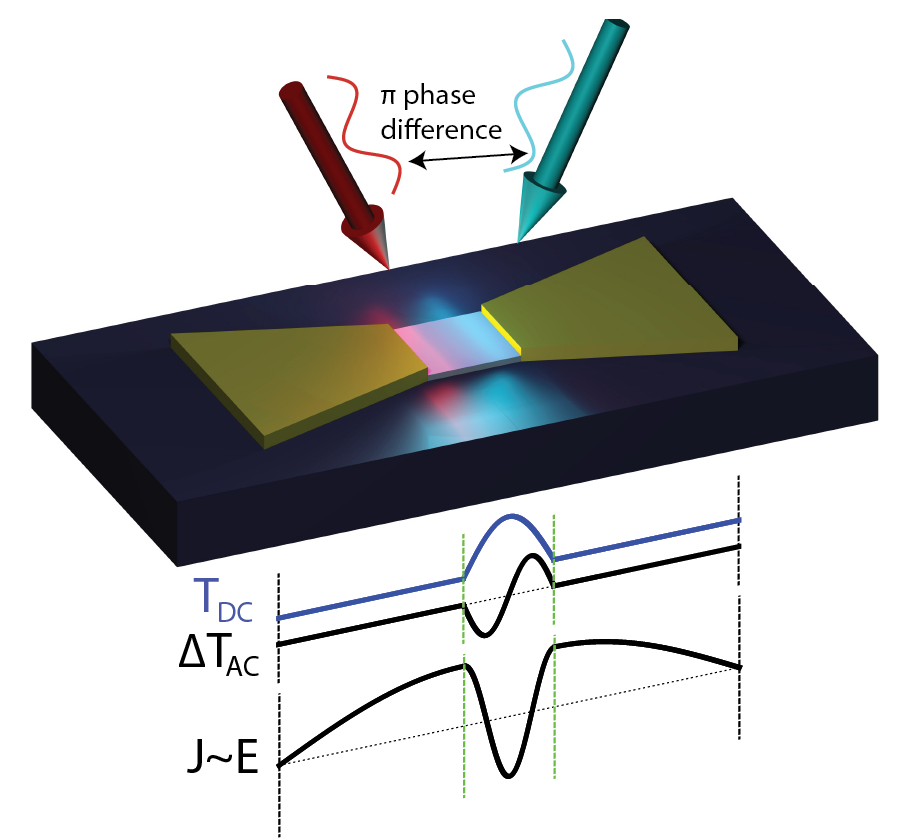}
\caption{\label{fig2}(color online) Two separate laser spots are centered on the two contacts to the graphene element, each consisting of two collinear NIR beams with a difference frequency in the THz spectral region, are temporally out of phase by $\pi$  radians and generate a dipolar plasma excitation that can be coupled to the far field using a symmetric dipole antenna.  This version of the emitter requires higher spatial resolution for the NIR illumination.}
\end{figure}

Therefore, Eq. \ref{eqCurrentAmplitude1} can be written as
\begin{equation}\label{eqCurrentAmplitude2}
J_0= \frac{\frac{2}{\pi} e \frac{\sigma'}{\sigma} k v_F^2 \omega (\omega + i \gamma) I}{[\omega (\omega + i \gamma) - \omega_2^2] [\omega (\omega + i \gamma) - \omega_p^2]-[(k_B T)\frac{\sigma'}{\sigma} \omega_2 \omega_p)]^2}
\end{equation}
where $\omega_p$ is the plasma resonance frequency given by $\omega_p^2 = \frac{n e^2 }{2 m \bar{\epsilon}} \frac{\pi}{L}=\frac{E_F e^2}{2\hbar^2 \bar{\epsilon} L}$ and $\omega_2^2 = (k v_F)^2 / 2$ which corresponds to second sound --- an entropy wave traveling close to the Fermi velocity.  The resonant form of Eq. \ref{eqCurrentAmplitude2}  suggests plasmons coupled to second sound.  Such a possibility has been recently discussed.\cite{11-phan2013} However, it is expected that the second sound mode is strongly Landau damped and that it “morphs” into the plasmon for the finite doping case we are considering. \cite{11-phan2013} Indeed the frequency terms in the numerator of Eq. \ref{eqCurrentAmplitude2} greatly diminish the second sound contribution compared with the plasmon since $\omega_2<\omega_p$ so that this semiclassical treatment may be consistent with that of ref. \citenum{11-phan2013}.

The resulting power delivered to the graphene plasma resonance from the NIR source is then:

\begin{align}\label{eqPowerEmitter}
\begin{split}
P_\mathrm{in} &= \frac{1}{2} W \, \mathrm{Re}\bigg[\int_0^L \! J^\ast \cdot E \, \mathrm{d}x \bigg] = \, \mathrm{Re}\bigg[\frac{W}{4 \sigma}\bigg]|J_0|^2 \\
 & = \frac{W \pi \hbar^2}{4 E_F e^2}|J_0|^2
 \end{split}
\end{align}	
			
To estimate the expected output of this emitter we use Eq. \ref{eqPowerEmitter} to calculate the power in the plasmonic mode.  This also gives an estimate of the expected radiation output of the device, since coupling the emitter element properly to an antenna would allow the emission of approximately half of this power as narrow band THz radiation in the far field. Since the device will not function well when the temperature rise is high, due to increased conduction to the lattice and reduced carrier mobility, it is useful to express the output in terms of the maximum DC temperature rise at the center of the graphene strip, $(T-T_0)_\mathrm{max}=4 I \Lambda \sigma(0)/L^2 \approx 150 \, \mathrm{K}$.  This gives rise to a strong dependence on $L$. We take $(T-T_0)_\mathrm{max} \approx 150 \, \mathrm{K}$ for the purpose of these estimations. The output is also strongly dependent on the mobility of the graphene.

As an example, for $L = 0.5\,\mu\mathrm{m}$, $W=10\,\mu\mathrm{m}$, $E_F = 40\, \mathrm{meV}$ , $\mu = 10,000 \,\mathrm{cm}^2 \mathrm{V}^{-1} \mathrm{s}^{-1}$, $\sigma(0)\approx 2\times 10^{-4} \, \Omega$, and a temperature rise of $150\,\mathrm{K}$  corresponding to an input NIR intensity of $10^4 \,\mathrm{W}/\mathrm{cm}^2$, the resulting $P_\mathrm{in}$ is $4 \,\mu \mathrm{W}$ at the plasma resonance at $2.4 \, \mathrm{THz}$ (the relative dielectric $K\equiv \bar{\epsilon}/\epsilon_0=7$  was taken to correspond with graphene on SiC or Si).  Increasing the mobility to $\mu=100,000\,\mathrm{cm}^2\mathrm{V}^{-1}\mathrm{s}^{-1}$ leads to a $P_\mathrm{in}$ comparable to the incident NIR power of approximately $1\,\mathrm{mW}$.  However, when the $\frac{1}{2}\mathrm{Re}(E\cdot J^\ast)$ term of Eq. \ref{TransportEq} is taken into account, we expect that the output power will be limited by the Carnot efficiency of the device as well as the coupling efficiency to the antennas.

Taking an effective relative dielectric of $K=3$, corresponding to graphene on SiO$_2$, we obtain an output $P_\mathrm{in}\approx 2 \,\mu\mathrm{W}$ at the plasma resonance at $4\,\mathrm{THz}$.  Again, increasing the mobility to $\mu=100,000\,\mathrm{cm}^2\mathrm{V}^{-1}\mathrm{s}^{-1}$  leads to $P_\mathrm{in}$ comparable to the incident NIR power.  We see, therefore, that the device promises microWatt output power over the THz range. How high in frequency this device can be expected to operate depends on the assumption of quasi thermal equilibrium, which is established through the rapid electron-electron scattering.  Therefore the upper limit over which this thermo-electric mechanism can be expected to operate is given by $\omega \tau_{e-e}<1$. For $\tau_{e-e}\sim 10 \,\mathrm{fs}$ this implies an upper limit of order $10\,\mathrm{THz}$.

Zomer et al.\cite{10-zomer2011} have achieved $\mu=125,000\,\mathrm{cm}^2\mathrm{V}^{-1}\mathrm{s}^{-1}$ for $n=4.3\times 10^{10} \mathrm{cm}^{-2}$  at $300\,\mathrm{K}$ on an h-BN substrate.  This corresponds to an $E_F$ of only $24\,\mathrm{meV}$. The resulting resistivity is $\sim 10^3 \, \Omega$ which impacts the process of coupling the power to the antenna as discussed below.  Also we note that the theory of supercollision cooling predicts that $\xi\propto\gamma^{-1/2}$ and otherwise independent of $E_F$.\cite{7-song2012}  Thus, for higher mobility graphene, domination by diffusion is more easily satisfied.

Many other configurations are possible each with their own advantages and disadvantages. For example, by using a more complex excitation scheme, the half-wave delay may be eliminated. Figure \ref{fig2} illustrates one such arrangement. Two laser spots temporally out of phase by $\pi$ radians heat the sample at different points. Thus, one side of the graphene is heated while the other cools (relative to the mean) resulting in a dipole mode that can be coupled directly to a conventional symmetric antenna.   This arrangement requires that the wavelength of the lasers, which are used for heating, are smaller than that of the plasmon wavelength allowing the small pattern to be generated. It also requires careful focusing and positioning as shown in the figure. Other possibilities include longer graphene segments which can support multiple periods of standing plasmons. This allows more driving power to be applied with the potential for more output power. Longer segments may actually benefit from heat conduction to the supporting substrate. Another consideration is the dielectric constant of the supporting substrate, which can allow control of the spatial frequency of the plasmon resonance. Also,  the output of the emitter scales directly with the width of the graphene. This allows more laser power to be directed onto the graphene and a lowering of the matching impedance.

\section*{Photo-thermal-electric p-n junction emitters}
The configurations shown in Figures \ref{fig1} and \ref{fig2} excite plasma resonances and therefore are particularly effective as emitters at relatively high THz frequencies as can be seen from Eq. \ref{eqCurrentAmplitude2}. However, they do not make good use of the Drude band of the optical response of graphene and this device will not function well at low frequencies where the plasmon wavelength dictates a large graphene length L and therefore a high DC temperature rise.

\begin{figure}
\includegraphics[scale=.5]{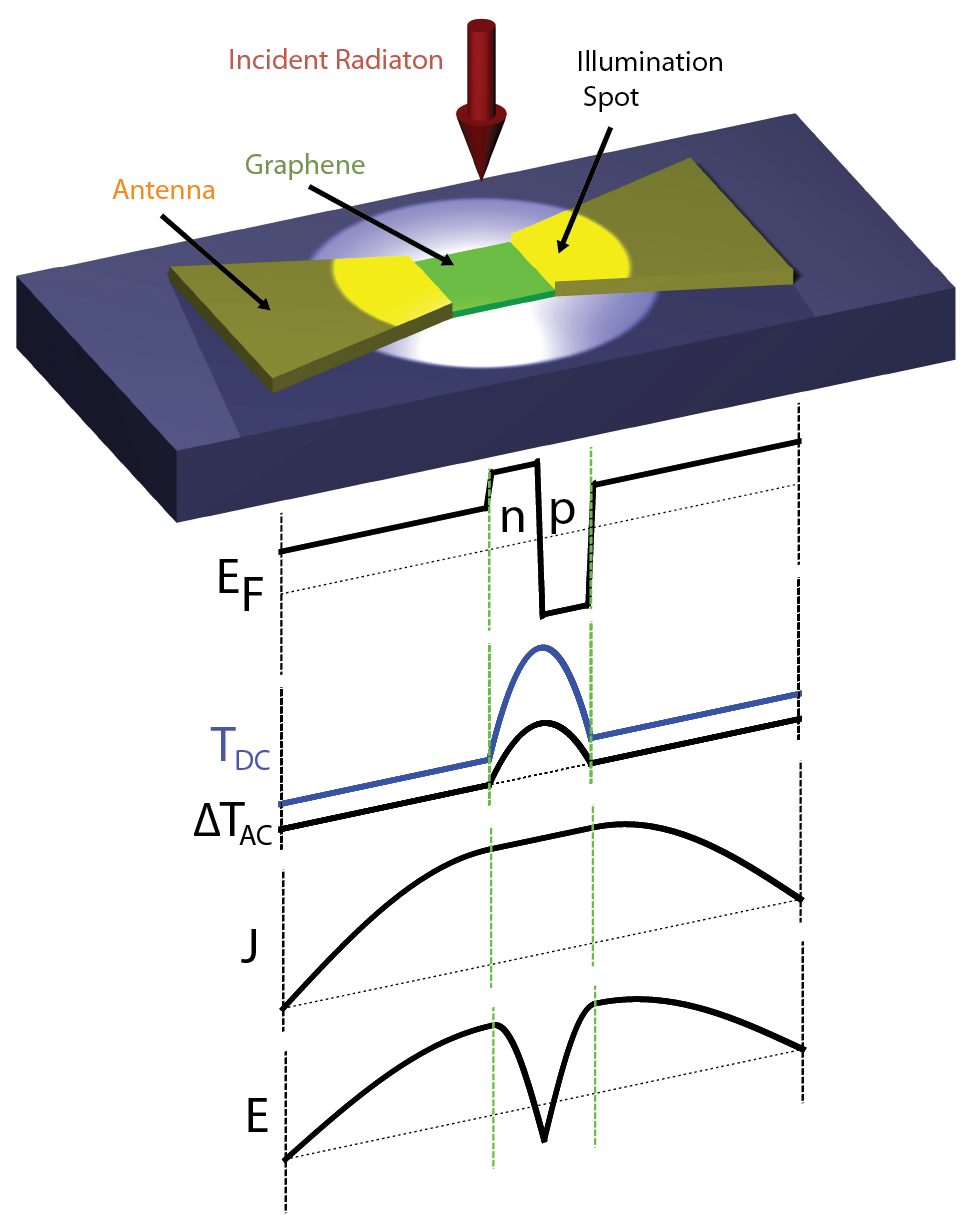}
\caption{\label{fig3}(color online) A graphene p-n junction device is uniformly illuminated by two interfering collinear laser beams with a THz difference frequency.  The NIR generated temperature profiles produce oppositely flowing electron currents in the n-region and hole currents in the p-region. In the low frequency limit, a uniform oscillating electrical current is produced that will couple efficiently to a dipole antenna. The spatial variations of the Fermi energy, temperature, electric current, and E-field are illustrated.}
\end{figure}

At lower frequencies a more efficient  configuration utilizes p-n junctions in graphene as illustrated in Figure \ref{fig3}. In this case the NIR generated temperature profiles produce oppositely flowing electron currents in the n-region and hole currents in the p-region resulting in a uniform oscillating electrical current that will couple efficiently to a dipole antenna. Consider the mode in which the currents on the two sides are in phase.  Here we assume both equal carrier densities and Fermi energies in the n and p regions.  There is a solution of the transport equations for the p-n junction in which $J=J_0 e^{-i \omega t}$, where $J_0$ is uniform, and the electric field is $E_0= \frac{e T \Lambda (\sigma'/\sigma) I_0}{\Lambda T \sigma (1-\beta)k^2 - i \omega C_e}|\sin \pi x / L |$, where $\beta = e^2 T^2 \Lambda (\sigma'/\sigma)^2$.  Consequently, there is a voltage drop across the device of $\Delta V = -2 \int_0^{L/2} \!  E_0(x) \, \mathrm{d}x$ and as a result a current satisfying $\Delta V = I_e (R+Z_L)$, where $I_e= W J_0$, $R(\omega) = L/W\sigma(\omega)$ is the impedance of the graphene element, and $Z_L$ is the impedance of the external circuit or antenna.  The power in the device is then determined by
\begin{equation}\label{eqPowerEmitter2}
\begin{split}
P = & \frac{\pi^2 \hbar^2}{2}  \bigg(\frac{v_F^2}{e E_F^2}\bigg)^2 \times \\ &\frac{I_0^2 (\sigma'/\sigma)^2 (\gamma^2 + \omega^2)}{|(1-\beta)\omega_2^2 - \omega (\omega + i \gamma)|^2} \mathrm{Re}(R(\omega)+Z_L)^{-1}
\end{split}
\end{equation}

An upper limit of the power available from the device is given for $Z_L=0$, so that $\mathrm{Re}(R(\omega)+Z_L)^{-1}=1/R(0)$, which is the DC conductance of the device.  Since $\beta\approx 1$ for typical operating conditions, the resonance frequency is less than the second sound frequency.  This low frequency resonance is in the single particle continuum ($\omega \leq k v_F$) and so should be broadened by Landau damping, which should flatten the low frequency response.  From Eq. \ref{eqPowerEmitter2} it follows that the output falls off as $\omega^2$.  The available power at $1\, \mathrm{THz}$ is estimated to be $50\,\mu\mathrm{W}$ for $L=0.3\,\mu\mathrm{W}$, $E_F=0.1\,\mathrm{eV}$, and $\tau = 1\,\mathrm{ps}$.

\section*{Discussion}

In general it should be possible to couple out an appreciable fraction of the power in these devices into radiation by employing appropriate antennas.  The p-n junction device is simply a dipole oscillator and will couple to a simple dipole antenna. However, all these devices require short lengths to limit the temperature.  For the plasmonic oscillators of Figures \ref{fig1} and \ref{fig2}, coupling the laser generated plasmonic energy into THz radiation will require antennas since the plasmonic wavelengths in graphene $\lambda_p$ are much smaller than free space wavelengths $\lambda_0$. At the plasma resonance the graphene presents a purely resistive load to the antenna as can be seen by considering the graphene element as a transmission line of length $\lambda_p/2$.  Therefore the fraction of the power that can be coupled to radiation is $P_a/P_\mathrm{total} \sim R_a/(R_a+R_g)$ where $R_a$ is the radiation resistance of the antenna and $R_g$ is the resistance of the graphene element. When matched, half of the power is convey to the antenna. Since the resistance of graphene is $30 \, \Omega_\square \lesssim R_\square \lesssim 5000 \, \Omega_\square$, matching is feasible and facilitated by adjusting the width $W$ of the graphene element. For the uniformly illuminated device of Fig. \ref{fig1}, the $n=0$ plasmon is monopolar and couples resistively with the antenna but with antiphase currents, which do not radiate well when coupled to a symmetric antenna.  To couple well to far field radiation for this case it is necessary to use an asymmetric antenna as illustrated in Fig. \ref{fig1}.  By having one arm approximately $\lambda_0/4$ long and the other $3\lambda_0/4$ long, similar to that of the offset, single-feed Windom antenna, the device will radiate efficiently.\cite{13-everitt1929} Essentially, the extra $\lambda/2$ section acts as a half wave delay. Alternatively a single quarter wave one sided antenna can be used but likely with lower efficiency.

One of the important applications of this emitter is as a local THz  oscillator.  Uses would include heterodyne detection of the emission lines of molecules at THz frequencies.  Surprisingly, even though the emitter uses a thermal process to generate the plasmons, the spectral width of the output is limited only by the width of the exciting lasers. However, to insure that higher power lasers possess linewidths much narrower than the difference frequency desired it may be necessary to cool higher power junction lasers to reduce the  thermal broadening.  The heated graphene element only introduces the noise equivalent to the thermal black body, which is only $k_B T$ per mode. For a graphene temperature as high as $1000\,\mathrm{K}$ this amounts to less than a picoWatt of power over the $50\,\mathrm{MHz}$ bandwidth of molecular emission lines.

\section*{Conclusion}

In summary, we have described devices for THz emission based on mixing of two NIR laser signals on graphene elements through the photo-thermo-electric effect.  The emission is monochromatic with a bandwidth determined by that of the NIR laser sources.  Plasmonic resonances provide a resonant response for uniformly doped graphene at high THz frequencies.  For p-n junctions of graphene, low THz frequency oscillators are expected.  The resulting THz power may be coupled to the far field by antennas.

\section*{Acknowledgements}
 We acknowledge useful discussions with Michael Fuhrer, Euyheon Hwang, Thomas Murphy, Justin Song, Andrei Sushkov, and Victor Yakovenko.  This work was supported in part by ONR grant \# N000141310865.

\bibliography{GrapheneEmitterBib}

\end{document}